\documentclass[twocolumn]{revtex4}
\usepackage[dvips]{graphicx}
\usepackage{amssymb,amsfonts,amsmath}

\newlength{\widthadd}  
\newlength{\heightadd}  
\begin{document}

\title{Survival benefits in mimicry: 
 a quantitative framework}

\author{Alexey Mikaberidze and Masudul Haque}

\affiliation{Max Planck Institute for the Physics of Complex Systems,
Noethnitzer Strasse 38,  01187 Dresden, Germany}

\begin{abstract}

Mimicry is a resemblance between species that benefits at least one of the
species.  It is a ubiquitous evolutionary phenomenon particularly common among
prey species, in which case the advantage involves better protection from predation.
We formulate a mathematical description of mimicry among prey species, to
investigate benefits and disadvantages of mimicry.  The basic setup involves
differential equations for quantities representing predator behavior, namely,
the probabilities for attacking prey at the next encounter.
Using this framework, we present new quantitative results, and also provide a
unified description of a significant fraction of the quantitative mimicry
literature.  The new results include `temporary' mutualism between prey
species, and an optimal density at which the survival benefit is greatest for
the mimic.
The formalism leads naturally to extensions in several directions, such as the
evolution of mimicry, the interplay of mimicry with population dynamics, etc.
We demonstrate this extensibility by presenting some explorations on
spatiotemporal pattern dynamics.

\end{abstract}

\maketitle

\section{Introduction}

Similarity between coexisting prey species is widely observed in nature.
A less defended prey species, \emph{e.g.}, a less poisonous or more palatable
species, may gain survival advantage by resembling a better defended prey
species.  Following the literature, we will refer to the lesser defended
species as `mimic' and the better defended one as `model'.
Mimicry, being a prime example of evolution by natural selection, is a key
topic in evolutionary theory and has been analyzed as such since Darwin's
time; some historical perspective is provided in Refs.\
\cite{JoronMallet_1998, MalletJoron_AnnRevEcolSyst99,
Mallet_commentary_PNAS01, Sherratt_evolution-review2008}.
Mimicry is also the subject of a growing number of experimental studies
\cite{Speed-etal_wildbirds_ProcRSoc00, Ihalainen-etal_JornEvBiol07,
Rowland-et-al_Nature07, BarberConner_acousticmimicry_PNAS07,
Ihalainen-et-al_BehavEcoSociobiol08}.
Despite its importance in evolutionary theory, mimicry has not received
extensive treatment in the mathematical biology literature.  In this article,
we formulate a continuum framework describing the basic dynamics of mimicry
and predation, which can serve as the foundation for a more quantitative
direction of mimicry investigations.

The basic phenomenon of mimicry immediately raises a number of intriguing
questions.  Can the presence of a lesser-defended mimic species actually be
advantageous to the unpalatable model?  More broadly, what are the conditions
for mimicry to be \emph{mutualistic} rather than \emph{parasitic}?  Given
fixed (un)palatabilities of each prey species and a certain population density
of the model, what population density would provide optimal protection for the
mimic species?  These questions are already nontrivial in the simplified
situation with two prey and one predator species, even in the absence of
population dynamics, spatial variations, etc.  In its full glory, mimicry
dynamics presents a vast array of fascinating questions and phenomena.  Only a
small fraction of this promising field has been considered quantitatively.

Mathematical modeling of mimicry benefits appears already in M\"{u}ller's
original nineteenth-century work (\emph{cf.}\ Refs.\ \cite{JoronMallet_1998,
Sherratt_evolution-review2008}).  More recently, some quantitative questions
have been addressed mainly through computer simulation \cite{OwenOwen84,
TurnerKearneyExton1984, huheey_AmNat88, speed93, TurnerSpeed_simulations96,
TurnerSpeed_review99, Speed_RobotPredators_AnimBehav99,
SpeedTurner_spectrum99, Darst_AnimBehav06}.  Nevertheless, there is a relative
shortage of mathematically well-defined questions and fully quantitative
treatments, which we aim to address in this work.

\begin{figure}
\centerline{\includegraphics[width=0.98\columnwidth]{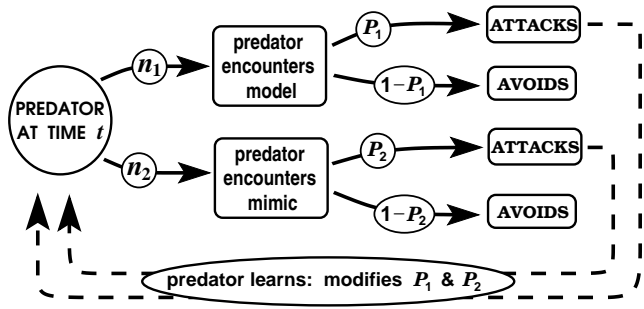}}
\caption{Predation and learning processes.  The predator encounters prey at
  rates $n_1$ and $n_2$, and attacks them according to the probabilities $P_1$
  and $P_2$ that characterize its behavior.  Each attack event leads to
  learning, \emph{i.e.}, a modification of $P_1$ and $P_2$.  
}\label{fig_setup}
\end{figure}

We focus first on the simplest possible case, namely, two prey species and a
predator species, with unlimited populations and constant densities $n_1$ and
$n_2$ that are not depleted by predation.  The predator attacks prey (model
and mimic) at each encounter with probabilities $P_1$ and $P_2$.  Predator
learning is encoded via the modifications of $P_1$ and $P_2$ due to each attack event.
The basic dynamics is illustrated schematically in Figure \ref{fig_setup}.
Since the simplest version does not include population dynamics, the only
dynamical variables are the attack probabilities $P_1$ and $P_2$.  
In computer simulations, each predator (in a predator ensemble) is associated
with its own $P_1$, $P_2$ values.  These values are modified via certain rules
each time that predator attacks a prey.
For our formulation, we also assume a large enough number of predators
such that only average $P_1(t)$, $P_2(t)$ values are relevant.  This allows us
to write down differential equations for these dynamical variables.
The resulting continuum description provides a context in which questions can
be posed and answered with mathematical rigor.  In addition to the questions
we examine here for the simplified scenario, the framework also allows easy
extension to more complex cases, such as varying population densities, spatial
inhomogeneities and pattern formation, additional species, etc.

%We focus in this article on the simplified scenario (Figure \ref{fig_setup}).
%
In several important cases, our differential equations allow exact analytic
expressions for the time dependent functions $P_1(t)$, $P_2(t)$.  These
solutions reveal new effects, such as (1) transient answers to the
mutualistic/parasitic question, which have strong implications for the
interpretation of mimicry experiments; (2) an optimum density of palatable
mimics at which mimicry most effectively provides survival benefits to the
mimic.
Our exact solutions also place on a firm mathematical footing and provide
clearer understanding of some existing computer simulation results, such as
the role of predator memory.
% in allowing mutualistic outcomes for cases of unequally prey defense.
%
%% As a further application, we seek an example case of a palatable species being
%% advantageous to the model, and show explicitly that this never occurs within a
%% relatively wide class of predator behavior.
%
In addition, we exploit the easy extensibility of our formulation to
incorporate simple spatial dynamics and explore spatiotemporal density
patterns --- we report a ``transmission'' effect of spatial patterns between
prey densities.
Some further applications are presented in the appendix.

\section{The mathematical framework}

The probability for a predator to encounter a model or mimic species is
determined by their densities $n_\mathrm{1}$ and $n_\mathrm{2}$.  We choose
units such that $n_\mathrm{1}$, $n_\mathrm{2}$ are the encounter probabilities
per unit time.  Our dynamical variables are $P_i(t)$, the attack probabilities
at each encounter.  (The index $i$ runs over values 1 for the model and 2 for the
mimic.)

The degree of defense for the model and mimic is characterized by
``palatabilities'' $\lambda_{1}$ and $\lambda_{2}$, which are the asymptotic
attack probabilities for an infinite-memory predator trained by an infinite
number of attack events.  A larger $\lambda_i$ indicates a more palatable
prey.  As the mimic species emulates a more defended model species, we will
use $\lambda_1\leq\lambda_2$.  Since $\lambda_i$ are probabilities, they are
restricted to lie between 0 and 1.  We will use $P_{1}(0)=P_{2}(0)=P_0=0.5$
for the initial (na\"ive or ``untrained'') attack probabilities.

Predator learning is modeled through the influence of attack events on $P_1$
and $P_2$ (figure \ref{fig_setup}).  After a predator attacks a model, its
$P_{1}$ value moves toward its asymptotic value $\lambda_1$, by an amount
determined by the learning coefficient $\alpha$:
\begin{equation} \label{eq_P1-after-model-attack}
P_{1} \xrightarrow{\rm model\ attacked} P_{1} + \alpha [ \lambda_{1}
-P_{1} ],
\end{equation}
Since the predator cannot perfectly distinguish between model and mimic, the
model attack probability also changes after a mimic attack event:
\begin{equation} \label{eq_P1-after-mimic-attack}
P_{1} \xrightarrow{\rm mimic\ attacked} P_{1} + r \alpha [ \lambda_{2} -P_{1} ].
\end{equation}
Here $r$ is a resemblance coefficient representing the quality of mimicry;
$r=0$ means no resemblance and $r=1$ means perfect resemblance. 
Since the average number of models attacked per unit time is $n_1P_1$ (figure
\ref{fig_setup}), the model attack event of equation
\ref{eq_P1-after-model-attack} contributes a term $n_1P_1 \times
\alpha(\lambda_1-P_1)$ to the rate of change of $P_1$.  With a similar
contribution from equation \ref{eq_P1-after-mimic-attack}, we obtain
\begin{equation} \label{eq:diff-eq-prob-forg1}
\frac{dP_1}{dt}= \alpha n_1 P_1 [ \lambda_1 -P_1 ] + r \alpha n_2
P_2 [\lambda_2 - P_1] + \gamma [P_0-P_1] \, .
\end{equation}
In addition to the predator learning in equations
(\ref{eq_P1-after-model-attack},\ref{eq_P1-after-mimic-attack}), we have also
included a ``forgetting'' term, quantified by a forgetting parameter
$\gamma$. In the absence of learning events, the attack probability decays to
the na\"ive value $P_0 = 0.5$.
Similarly, for the mimic attack probability:
\begin{equation}\label{eq:diff-eq-prob-forg2}
\frac{dP_2}{dt}= \alpha n_2 P_2 [ \lambda_2 -P_2 ] + r \alpha n_1
P_1 [\lambda_1 - P_2] + \gamma [P_0-P_2] \, .
\end{equation}

The analysis presented in this article is based on the coupled nonlinear
equations (\ref{eq:diff-eq-prob-forg1},\ref{eq:diff-eq-prob-forg2}) and
variations or extensions thereof.
Unless otherwise specified, we will generally be using $\alpha=1$,
$n_1=n_2=0.5$.

The leading terms of (\ref{eq:diff-eq-prob-forg1},\ref{eq:diff-eq-prob-forg2})
have the form of logistic equations well-known in population dynamics.
However, our equations are for probabilities representing predator
decision-making (not for populations or densities) and are based on quite
different considerations.

In addition to the attack probabilities, another informative quantity in
judging the benefits/losses of mimicry is the prey mortality, i.e. the number
of prey that have been attacked by the time $t$:
\begin{equation}\label{eq:n-killed}
N_i(t) = \int_{0}^{t} n_i P_i(t') dt'  \; ,   \quad  \quad  [i=1,2] \; .
\end{equation}
%
%% Here and in the following equations the index $i$ runs through values 1 for
%% model and 2 for mimic quantities. 
%
This quantity is particularly relevant in making contact with experimental
data where mimicry advantages are usually reported in terms of prey
mortalities \cite{Speed-etal_wildbirds_ProcRSoc00, Rowland-et-al_Nature07}.
To quantify mimicry benefits, we will more often use the ``favorability''  ratio
\begin{equation}
f_i^{(r)}(t) ~=~ P_i^{(r=0)}(t) \;  / \; P_i^{(r)}(t) \; ,
\end{equation}
especially its asymptotic value $f_i(t\rightarrow\infty)$.  
This ratio compares attack rates in non-resembling ($r=0$) and resembling
($r>0$) cases
Mimicry is beneficial to model (mimic) if $f_1>1$ ($f_2>1$).

\begin{figure}
\begin{center}
\centering
\includegraphics*[width=0.98\columnwidth]{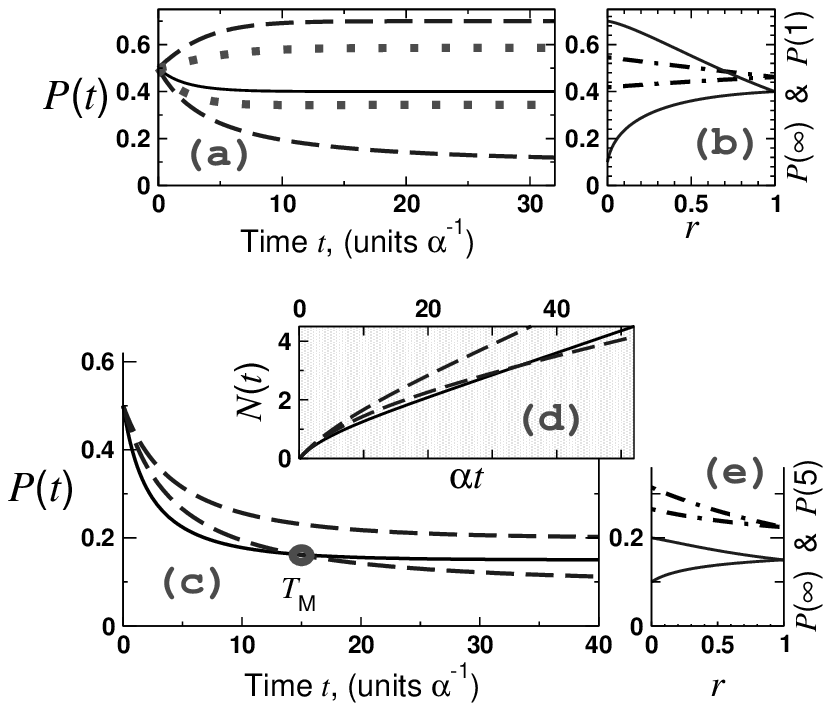}
\caption{ Results for infinite memory ($\gamma=0$).
\\ \textbf{(a,b)} Attack probabilities for palatable mimic, $\lambda_1= 0.1$,
$\lambda_2= 0.7$.
\\ \textbf{(a)} Dashed, dotted: $r=0, 0.4$.  In each pair, lower (upper) curve
represents model (mimic).  Solid curve is $r=1$ probability, common for mimic
and model.
\\ \textbf{(b)} Dependence on resemblance.  Dash-dot: $\alpha{t}=1$, solid:
asymptotic.  Model and mimic curves merge at $r=1$.  
\\ \textbf{(c-e)} Unpalatable mimic, $\lambda_1= 0.1$, $\lambda_2= 0.4$.  
\\ In \textbf{(c,d)}, dashed pair are $r=0$ curves (upper: mimic, lower: model),
and solid curves are common $r=1$ quantities.
\\ \textbf{(c)} Attack probabilities.  The solid $r=1$ curve crosses the lower
dashed (model) $r=0$ curve at time $T_{\rm M}$ up to which mutualism persists.
Mutualism corresponds to solid $r=1$ curve being lower than both dashed $r=0$
curves.
\\ \textbf{(d)} Mortalities.  Also displays the same transient phenomenon via a
crossing of $r=1$ and model $r=0$ curves.
\\ \textbf{(e)} Dependence on resemblance.  Dash-dot: $\alpha t = 5$, solid:
asymptotic.  The negative slope of the model (lower dash-dot) curve at finite
time indicates the transient mutualism.
}\label{fig_infinite-memory}
\end{center}
\end{figure}

\section{Context: assumptions and relevance}

\subsection{Predator psychology}

Obviously, the equations
(\ref{eq:diff-eq-prob-forg1},\ref{eq:diff-eq-prob-forg2}) make strong
assumptions about the predation process.
Learning and forgetting of avoidance are quantified through simple and
reasonable rules.  However, other rules could be argued to be equally
reasonable (\emph{c.f.}, \cite{TurnerSpeed_review99, Speed_EvolEcology99,
Sherratt_evolution-review2008} for detailed discussions).  One alternative
point of view \cite{TurnerKearneyExton1984, JoronMallet_1998,
MalletJoron_AnnRevEcolSyst99} is that the asymptotic attack probability for
defended prey should be always zero, and that the degree of defense
(palatability) should be quantified only by differences in the learning rates
$\alpha$.  M\"uller's original analysis also implicitly uses this picture of
learning \cite{JoronMallet_1998, Sherratt_evolution-review2008}. Such a
situation is conceivable when the predator has ample alternative sources of
nourishment.
Although our equations can be easily modified to incorporate such a picture of
learning, we will here restrict ourselves to the case of nonzero asymptotic
attack probabilities.  This is supported by experiment
\cite{Speed_EvolEcology99}, and presumably reflects the common situation where
predator species are not fully saturated through alternate prey.

Predator behavior is an active topic of research and is not yet fully
understood or characterized (\emph{c.f.}, Refs.~\cite{Darst_AnimBehav06,
Lynn_AnimBehav05,SkelhornRowe_AnimBehav06, Sherratt_evolution-review2008}).
Given the present level of understanding, it is reasonable to use the simplest
way of quantifying predator training.
Accordingly, we have mostly left out palatability-dependences of the learning
rates, $\alpha(\lambda)=\alpha$, and also used the simplest description of
forgetting.  
In the appendix, we include some variations of
these simple assumptions when we seek to make generic statements, such as the
(non)existence of mutualism with palatable prey.
Refs.~\cite{TurnerSpeed_simulations96, Speed_RobotPredators_AnimBehav99,
SpeedTurner_spectrum99} describe various learning and forgetting rules used in
the literature.

\subsection{The palatability spectrum}

One extreme of mimicry is the case where the mimic lacks defenses,
\emph{i.e.}, is palatable, so that mimicry is parasitic with no benefit to the
model.  This is known as Batesian mimicry.  At the other extreme, the two prey
species could be equally defended, so that mimic and model are interchangeable
labels.  This is known as Muellerian mimicry.  More interesting cases involve
unequally defended prey, constituting a spectrum between the two extremes
\cite{SpeedTurner_spectrum99}.

%In our framework, palatability has a strict meaning: a prey is palatable
%(unpalatable) if $\lambda>P_0$ ($\lambda<P_0$).

\subsection{Benefits versus losses}

Parasitism versus mutualism is a major theme of theoretical ecology
(\emph{e.g.}, \cite{VandermeerGoldberg_book_PopulatnEcology03}).
In the mimicry context, the benefit/loss issue has been discussed with
computer simulations, but unfortunately has not received as thorough a
mathematical treatment as in other contexts.  A formulation like the one we
present in this Article is a necessary step in this direction.

\subsection{Evolutionary implications}

The issue of benefits versus losses in mimicry phenomena is important from an
evolutionary perspective; there are vital implications for warning color
diversity and polymorphism phenomena \cite{JoronMallet_1998,
MalletJoron_AnnRevEcolSyst99, Mallet_commentary_PNAS01,
Sherratt_evolution-review2008, Mallet_EvolEcology99, Servedio_Evolution00}.
Our setting is well suited for addressing the basic question of whether mimicry
is mutualistic or parasitic.
A full quantitative study of evolutionary implications requires a more
involved (perhaps multi-generation or many-species) framework.  In the
Supplementary Information we show how our formalism can be extended to address
some of these themes.

\subsection{Differential equation formulation}

Our formulation is a continuum one, in contrast to the individual-predator
simulations that have dominated theoretical mimicry studies.  Such a setup
opens up many new possibilities.  It affords exact solubility for several
important cases, even in the presence of forgetting.  The formulation allows
easy extension in several directions, such as spatial inhomogeneity, as we
report below.  It allows analytic tools, such as pattern formation theory, to
be applied directly to such extensions.
In addition, differential equations encourage us to study the complete
predator training process, \emph{i.e.}, to focus on the dynamical aspects of
mimicry.  This is in contrast to existing literature which focuses on
steady-state situations where the predator is already fully trained, which
corresponds to the asymptotic behavior of our solutions.

\begin{figure*}
\begin{center}
\centering
\includegraphics*[width=0.98\textwidth]{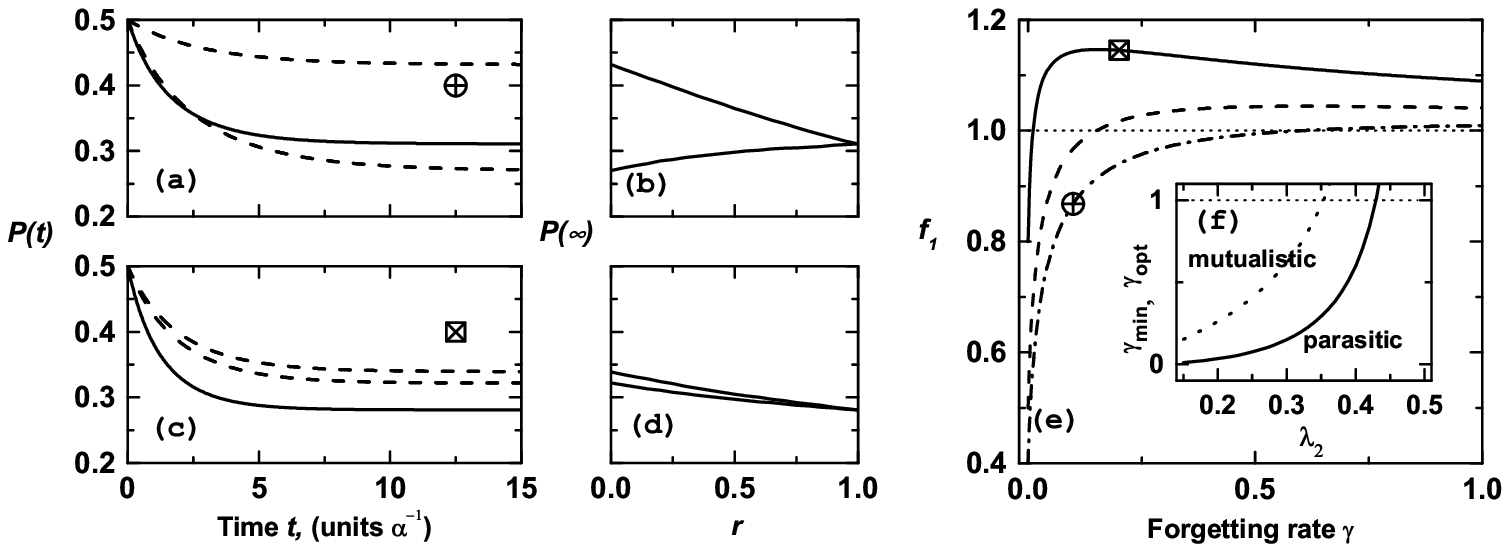}
\caption{
Finite memory ($\gamma\neq0$), $\lambda_1=0.1$. 
\\ \textbf{(a,b)} No asymptotic mutualism, $\lambda_2=0.4$, $\gamma=0.1$.
\textbf{(c,d)} Asymptotic mutualism, $\lambda_2=0.15$, $\gamma=0.2$.
\\ In \textbf{(a)} and \textbf{(c)} dashed lines show $r=0$ cases (lower - model, upper
- mimic); solid lines show $r=1$ probabilities common to model and mimic. 
Note that \textbf{(b,d)} show asymptotic attack rates $P(\infty)$ only.  The
positive and negative signs of $dP_1/dr$ [slopes of lower curves in
\textbf{(b)} and \textbf{(d)}] represent parasitism and mutualism
respectively.
\\ \textbf{(e)} Asymptotic model favorability. Solid, dashed, dash-dot:
$\lambda_2=0.15, 0.3, 0.4$.  Parameters corresponding to \textbf{(a,b)} and
\textbf{(c,d)} are marked with circle and square respectively.
\\ \textbf{(f)} Solid curve, $f_1(\gamma_{\rm min})=1$, separates mutualistic and and parasitic
mimicry, dotted line indicates parameters optimum for model where
$f_1(\gamma)$ has a maximum.
}\label{fig_forgetting}
\end{center}
\end{figure*}

\section{Example applications}

\subsection{Infinite predator memory}

We first report the dynamics of predator training with unlimited memory,
$\gamma=0$.
%
%% We first report the dynamics of the equations
%% (\ref{eq:diff-eq-prob-forg1},\ref{eq:diff-eq-prob-forg2}) without forgetting,
%% $\gamma=0$.
%
With no resemblance ($r=0$), equations
(\ref{eq:diff-eq-prob-forg1},\ref{eq:diff-eq-prob-forg2}) are not coupled and
reduce to the logistic equation:
\begin{equation}\label{eq:diff-eq-prob}
\frac{dP_i}{dt} ~=~ \alpha \, n_i  P_i \;  [ \lambda_i -P_i ] ,
\end{equation}
which yields exact solutions for attack probabilities and mortalities:

\begin{equation} \label{eq:prob-sol}
P_i(t) = \frac{\lambda_i P_{0}}{P_{0} - (P_{0}-\lambda_i) \exp(-\alpha n_i
  \lambda_i t)} \; ,
\end{equation}
\begin{equation} \label{eq:num-killed}
N_i(t) = \frac{1}{\alpha} \ln [P_0 (\exp (\alpha n_i \lambda_i t)
-1 ) + \lambda_i] - \ln \lambda_i / \alpha \; .
\end{equation}
The $P_i(t)$ here are logistic functions.  The prey mortality grows linearly at
small times ($N_i \sim{n_iP_0t}$) and also at asymptotically large times ($N_i
\sim n_i\lambda_i{t}$),
with mortality rates determined respectively by na\"ive and asymptotic attack
probabilities ($P_0$ and $\lambda_i$).

For perfect resemblance, $r=1$,
%
%the attack probabilities are the same, \emph{i.e.}, 
%
the solutions $P_1(t)$ and $P_2(t)$ are synchronized.  Equations
(\ref{eq:diff-eq-prob-forg1},\ref{eq:diff-eq-prob-forg2}) collapse to one
equation:
\begin{equation}\label{eq:diff-eq-probr1}
\frac{d P}{d t} = \alpha \, (n_1 + n_2)\,  P\; [ \bar{\lambda} -P ] \; ,
\end{equation}
where $P=P_1=P_2$, and $\bar{\lambda} = (n_1 \lambda_1+n_2 \lambda_2) /
(n_1+n_2)$.  The $P(t)$ and mortalities for $r=1$
are given by the same equations (\ref{eq:prob-sol},\ref{eq:num-killed}) as in
the $r=0$ case, with $n_i$ and $\lambda_i$ replaced by $(n_1+n_2)$ and
$\bar{\lambda}$ respectively.

%Describe Fig. 2

Figure \ref{fig_infinite-memory}(a,c,d) displays typical time-dependences for
attack probabilities and mortalities.  The $P_i(t)$ start at their na\"ive
value $P_0=0.5$ and tend to their asymptotic values, equal to $\lambda_1$ and
$\lambda_2$ for non-resembling prey and to $\bar{\lambda}$ for the case of
perfect resemblance.

For imperfect mimicry, $0<r<1$, the asymptotic 
%time-independent 
solutions can still be found analytically, as the fixed points of the
dynamical equations.  It is also easy to obtain the complete time dependence
numerically.  Attack probabilities and mortalities vary monotonically between
the perfect and zero resemblance cases described above (figure
\ref{fig_infinite-memory}b and \ref{fig_infinite-memory}e).
A negative slope of a $P_i$ versus $r$ curve indicates that mimicry is
beneficial for that prey species.

\subsection{Asymptotes and transients}

The infinite-memory equations predict that the presence of mimics is always
harmful for models at sufficiently large times -- the asymptotic model attack
probability for $r>0$ is always larger than that for $r=0$, \emph{i.e.}, 
$P_1^{(r>0)}(\infty)>P_1^{(r=0)}(\infty)$.  In other words, with this simplest
version of predator psychology, mimicry is always parasitic in the long run,
unless the prey species are equally defended ($\lambda_1=\lambda_2$).
% In this case mimicry is neither harmful, nor beneficial.

However, our exact solutions display a surprising \emph{transient} behavior,
namely, for unpalatable but less-defended mimics ($\lambda_1<\lambda_2<P_0$),
there is a finite time ($T_{\rm M}$) up to which mimicry can be favorable to
the model, as displayed in figure \ref{fig_infinite-memory}c-e.  In the
transient regime $t<T_{\rm M}$, the slope $dP_1/dr$ for the model is negative.
This transient mutualistic effect has not appeared previously in the
literature, and has significant implications for the interpretation of mimicry
experiments.

The appendix describes the extent of the transient regime
in more detail.

\subsection{The predators forget}

The inclusion of forgetting in the predator learning process has vital
consequences for mutualism in mimicry.
Some aspects, \emph{e.g.}, comparison of various memory mechanisms, have been
treated through computer simulations \cite{speed93, TurnerSpeed_simulations96,
SpeedTurner_spectrum99, Speed_RobotPredators_AnimBehav99}.
Our formalism gives exact equations that quantify effects of finite predator
memory, and reveals new effects.  In particular, we explore the question of
mutualism as a function of the rate of forgetting $\gamma$.  We find a
\emph{critical} value of the forgetting parameter above which mimicry becomes
asymptotically favorable to model.

We reinstate the memory parameter $\gamma$ in equations
(\ref{eq:diff-eq-prob-forg1},\ref{eq:diff-eq-prob-forg2}). With no resemblance
($r=0$), the decoupled equations have exact solutions
\begin{equation} \label{eq:no-resemblance-prob_forg}
P_i= \frac{1}{2 \alpha n_i} \Biggl( b_i + A_i \tanh \left[ \frac{A t}{2} -
\mathrm{atanh} \Bigl[ \frac{b_i - 2 \alpha n_i P_{0}}{A_i} \Bigr] \right] \Biggr) \; ,
\end{equation}
%
%% where
%% \begin{equation}
%% A_i = \sqrt {4 \gamma P_{0} n_i \alpha + {b_i}^2}  \; ;
%% \qquad  b_i = \alpha n_i \lambda_i- \gamma
%% \end{equation}
%
where $b_i = \alpha n_i \lambda_i- \gamma$ and $A_i = \sqrt {4 \gamma P_{0}
n_i \alpha + {b_i}^2}$.
The asymptotic value is
\begin{equation}
P_i(t\rightarrow\infty) = \lambda_i /2 + \frac{A_i - \gamma}{2 \alpha n_i}
\; .
\end{equation}
Not surprisingly, $\gamma$ shifts the asymptotic attack probability from
$\lambda_i$ toward the na\"ive value $P_0$.
As in the $\gamma=0$ case, the perfect resemblance ($r=1$) solution
($P=P_1=P_2$) can be obtained from the $r=0$ solutions above, via
$n_i\rightarrow(n_1+n_2)$ and $\lambda_i\rightarrow\bar{\lambda}$.
%
% We have not found analytic expressions for imperfect mimicry $0<r<1$.

The finite-memory learning process is illustrated by $P_i(t)$ curves in
figures \ref{fig_forgetting}a,c.  Figures \ref{fig_forgetting}a,b show a case
with only transient mutualism. Figures \ref{fig_forgetting}c and
\ref{fig_forgetting}d involve parameters (larger $\gamma$, smaller
$\lambda_2$) where mimicry is mutualistic at all times.
Figure \ref{fig_forgetting}e shows the asymptotic model favorability $f_1$ as
a function of the forgetting rate $\gamma$.  There is a threshold
$\gamma=\gamma_{\rm min}$ above which the mimicry is mutualistic ($f_1>1$),
and an optimum value at which $f_1>1$ is maximum (\ref{fig_forgetting}e and
\ref{fig_forgetting}f).

\begin{figure}
\centering 
\includegraphics*[width=0.98\columnwidth]{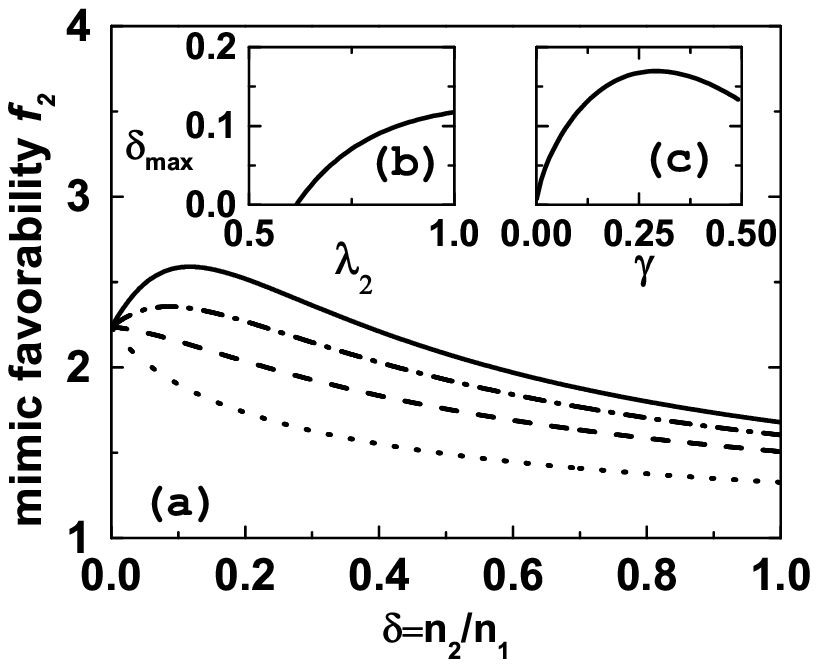}
\caption{ 
Non-monotonic dependence on mimic density.
\\ \textbf{(a)} Mimic favorability $f_2$ versus relative mimic density $\delta =
n_2/n_1$ (equation \ref{eq_fav2-vs-delta2}).  Here $\lambda_1=0.1$,
$\gamma=0.1$, $n_1+n_2=1$.  Dotted, dashed, dash-dot, solid: $\lambda_2= 0.3,
0.6, 0.8, 1.0$.  A maximum appears at large $\lambda_2$.
\\ Insets show the optimum mimic density as a function of 
\\ \textbf{(b)} model
palatability $\lambda_2$; \textbf{(c)} forgetting parameter $\gamma$.
}\label{fig_OwenOwen}
\end{figure}

\subsection{Optimum density}

We now consider variable prey densities.  
We have found that the benefit of resemblance to the mimic can depend
non-monotonically on the mimic density.
In particular, for palatable mimics, $\lambda_2>P_0$, there is an
\emph{optimum} density, at which the asymptotic mimic favorability has a
maximum.  Again, our formulation provides an explicit expression describing
this phenomenon:
\begin{equation}  \label{eq_fav2-vs-delta2}
f_2^{(r=1)}(\infty) = \frac{\lambda_2 /2 + 
(A_2 - \gamma)/2 \alpha n_2}{\bar{\lambda} /2 + (\bar{A} - \gamma)/[2 \alpha
    (n_1+n_2)]}  
\; .
\end{equation}
In figure \ref{fig_OwenOwen}a, we plot the favorability against the relative
mimic density $\delta = n_2/n_1$, where $n_1+n_2=1$ is fixed.  
The non-monotonic behavior is robust; the optimum density $\delta_{\rm max}$
has very weak dependence on the resemblance $r$.  The dependences of
$\delta_{\rm max}$ on $\lambda_2$ and $\gamma$ are shown in figure
\ref{fig_OwenOwen}b,c.

A prior example of non-monotonic density dependence \cite{OwenOwen84,
Speed_RobotPredators_AnimBehav99, MalletJoron_AnnRevEcolSyst99} required
unpalatable mimics ($\lambda_2<P_0$) and complicated forgetting rules.  For
example, the ``variable forgetting'' of Ref.\
\cite{Speed_RobotPredators_AnimBehav99} in our language would imply
density-dependent $\gamma$.  In contrast, we have found non-monotonic features
and optima with a simple and transparent description of forgetting.

Understanding the density-dependence of mimicry benefits is crucial for
studying mimicry together with population dynamics and evolution.  A more
detailed account of the density-dependence is provided in the appendix.

\subsection{Spatiotemporal dynamics and patterns}

\begin{figure}
\begin{center}
\centering
\includegraphics*[width=0.98\columnwidth]{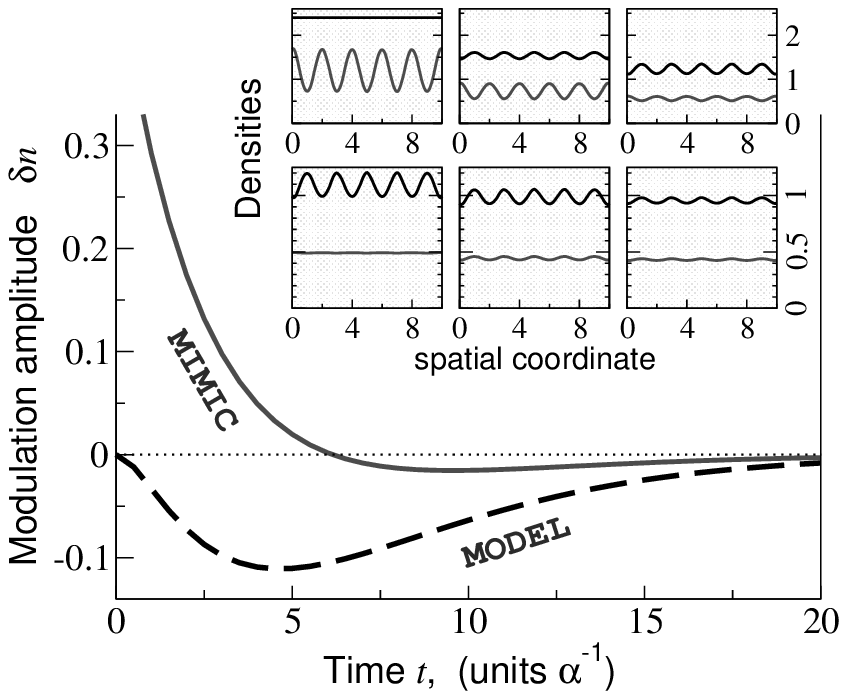}
\caption{Pattern transmission: an initial spatial modulation of the mimic
  density is transfered to the model density, and then transmitted back to the
  mimic density with opposite the original phase.
\\ The densities are of the form $n(x,t) = \bar{n}(t) + \delta{n}(t)\cos(kx)$.
Main curves show modulation amplitudes $\delta{n}$.  The sign of $\delta{n}$
indicates phase relative to the initial mimic modulation.
\\ Six insets show model (upper) and mimic (lower) density profiles, at times (top
row) $t= 0, 2, 4$ and (lower row, note different scale) $t=6, 10, 15$.
\\ (Parameters:  $r=0.9$, $\beta_1=\beta_2=1.0$,
$\bar{\beta_1}=0.3, \bar{\beta_2}=0.15$.)
}\label{fig_spatiotemp}
\end{center}
\end{figure}

One attraction of our formulation is the ease of extension to studying
spatiotemporal dynamics.
We have tried the simplest cases of one-dimensional spatial dependence, with
the densities $n_{i}(x,t)$ having growth and/or diffusion dynamics
compensating the losses due to predation.  This already supports a dramatic
``pattern transmission'' phenomenon, an example of which is shown in figure
\ref{fig_spatiotemp}.
Equations (\ref{eq:diff-eq-prob-forg1},\ref{eq:diff-eq-prob-forg2}) for
$P_{i}(x,t)$ (with $\gamma=0$) were supplemented by prey density equations:
\begin{gather}
\partial_t n_1(x,t) = -\beta_1 n_1(x,t) P_1(x,t) + \tilde{\beta}_1 
~+~ D_1\nabla^2n_1(x,t) \; , \\
\partial_t n_2(x,t) = -\beta_2 n_2(x,t) P_2(x,t) + \tilde{\beta}_2 
~+~ D_2\nabla^2n_2(x,t) \; .
\end{gather}
A constant growth rate ($\tilde{\beta}_i$) is not extremely realistic but
should be regarded as an effective description that provides a simple
mechanism for constant nonzero asymptotic densities.  The phenomenon we
emphasize (transfer of inhomogeneity from one prey density to the other) is
robust for a variety of growth and diffusion terms.  

We find that a spatial variation (\emph{e.g.}, modulation) in the initial
mimic density induces a spatial variation in model density, even when the two
densities are not directly coupled.  The predation ($P_{1}$, $P_{2}$ variables)
mediate a coupling between the two prey species.
The induced variation is ``out of phase'' for parasitic mimicry , i.e., a bump
in mimic density causes a dip in model density, and ``in-phase'' for
mutualistic cases.  Figure \ref{fig_spatiotemp} shows a parasitic example
($\lambda_1=0.1$, $\lambda_2=0.8$).  The $\bar{\beta}_i$ and $D_i$ terms
eventually smooth out all modulations.
In figure \ref{fig_spatiotemp}, after the mimic modulation has died out the
first time ($\alpha{t}\approx{5.3}$), the model density pattern in turn
induces a modulation in the mimic density.  Because mimicry is favorable to
the mimic species, the re-induced modulation is now in-phase, opposite to the
original mimic distribution.
Figure \ref{fig_spatiotemp} uses $D_i=0$; diffusion accelerates the
smoothening process but the transmission effect is not qualitatively affected.

For the parameters we explored, there was no spontaneous pattern
formation.  However, our setup, when extended in spatial dimensions and
additional dynamical processes, is clearly capable of a rich set of phenomena,
waiting to be explored.  In particular, equations
(\ref{eq:diff-eq-prob-forg1},\ref{eq:diff-eq-prob-forg2}) already contain
``reaction'' terms of the form $-P_1P_2$; it is therefore quite likely that
spatial diffusion can induce patterns in some geometries, via, \emph{e.g}, the Turing
reaction-diffusion mechanism.
Our framework provides the perfect setting for using the techniques of pattern
formation theory \cite{CrossHohenberg_PatternFormation_RMP93} to analyze such
phenomena.

\section{Conclusions}

The topic of mimicry has received less mathematical attention than could be
expected from its central importance to issues in evolutionary theory such as
polymorphism and biodiversity.
%
%% Despite being a key topic in evolutionary theory, mimicry has not received
%% extensive mathematical treatment.  
%
We have presented a framework which, although simple, has the advantage of
being amenable to rigorous analysis and to extensions in various directions.
In this regard, we recall the Lotka-Volterra equations in the mathematical
study of predator-prey number dynamics, which are by themselves too simple to
describe any ecological reality, but nevertheless provide the foundation for a
vast area of theoretical ecology, due to their mathematical simplicity and
rigor, and extensibility.  We hope that our work will similarly stimulate
quantitative developments in mimicry analysis.

Since the prime attraction of our formalism is its easy extensibility, we end
by discussing three promising directions.

First, as illustrated through one example in figure \ref{fig_spatiotemp}, the
inclusion of spatial dynamics opens fascinating pattern dynamics
possibilities.  Spatial distributions and mosaic structures in mimicry are
active topics of ecological research \cite{Ellner-etal_spatial_JourMathBiol98,
Sasaki-etal_mosaic_ThPopBiol02, JoronIwasac_spatial_JourTheorBiol05,
Sherratt_spatial_JourTheorBiol06, KawaguchiSasaki_spatial_JourTheorBiol06};
our formalism provides the groundwork for a unified mathematical treatment of
some of the many intriguing phenomena.

Second, through the dynamics of our resemblance ($r$) parameter, our setup
allows a basic description of \emph{evolutionary} dynamics.  (Some discussion
is provided in the appendix).  Incorporating more intricate
descriptions of the evolution of resemblance remains an important open
direction for future study.

Finally, with relatively small manipulations of our basic equations,
experiments on mimicry can be modeled quantitatively.  In this regard, the
dramatic \emph{transient} solutions we have provided may well turn out to be
vital.  For example, in a recent experiment \cite{Rowland-et-al_Nature07},
mutualistic behavior was found despite unequal prey defenses.  Within our
simple learning and forgetting rules, this can already have two different
explanations: \textbf{(a)} the reported mutualism could be due to the
transient effect described in figure \ref{fig_infinite-memory}c-e;
\textbf{(b)} the observed mutualism could be a true asymptotic effect like the
one found in the finite-memory cases, figure \ref{fig_forgetting}c-f.  Since
Ref.~\cite{Rowland-et-al_Nature07} records data after a fixed mortality, it is
not immediately obvious which period of the training dynamics the data
corresponds to.  However, the basic analysis of this article already provides
a framework within which such questions can be addressed.  The interpretation
of experimental data is thus one more promising application of our work.

\appendix

\section*{Appendix}

In the appendix we describe

\begin{enumerate}

\item the time up to which transient mutualism exists;

\item dependence on mimic density;

\item further applications of our formalism.

\end{enumerate}

\subsection{Transient mutualism}

\begin{figure}
\begin{center}
\centering
\includegraphics*[width=0.98\columnwidth]{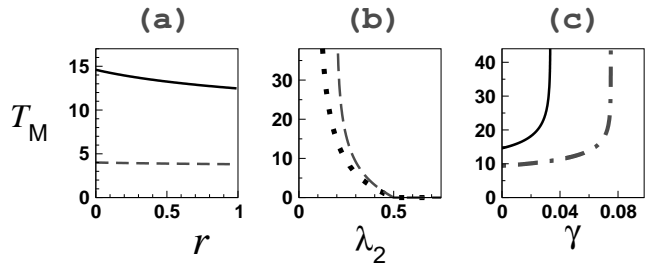}
\caption{ Time, $T_{\rm M}$, up to which transient mutualism persists. 
\\ \textbf{(a)} Plotted against resemblance paratmer $r$,  with $\gamma=0$ and
$\lambda_1=0.1$.  Solid: $\lambda_2=0.2$; dashed: $\lambda_2=0.35$.
\\ \textbf{(b)} Plotted against mimic palatibility $\lambda_2$, with $r=1$ and
$\gamma=0$.  Dotted: $\lambda_1=0.1$; dashed: $\lambda_1=0.2$.
The divergences show that for infinite predator memory asymptotic mutualism
only occurs for $\lambda_2=\lambda_1$.
\\ \textbf{(c)} Plotted against memory parameter $\gamma$, with $r=1$ and
$\lambda_1=0.1$.  Solid: $\lambda_2=0.2$; dash-dotted: $\lambda_2=0.25$.  The
divergences show that mutualism becomes asymptotic above a critical value of
the forgetting rate $\gamma$.
}\label{fig_transient-time}
\end{center}
\end{figure}

With infinite-memory predators, we have shown that although mimicry is not
mutualistic (favorable to both prey species) in the long run, there is a part
of the training period where temporary mutualism holds.
In figure \ref{fig_transient-time}, we explore this phenomenon further by
plotting the time ($T_{\rm M}$) up to which transient mutualism persists,
against various parameters.  The parameter $T_{\rm M}$ is defined pictorially
in figure 2 of the main text.

The plot against the resemblance parameter $r$ shows (figure
\ref{fig_transient-time}a) that the phenomenon is very robust.  The
transient time $T_{\rm M}$ changes little when the resemblance is varied from
perfect to almost non-existent.

In figure \ref{fig_transient-time}b, the mimic palatability is varied.  The
phenomenon only occurs for $P_0 > \lambda_2 > \lambda_1$.  The curve diverges
for $\lambda_2 \rightarrow \lambda_1$, which indicates that asymptotic
mutualism is possible for equally defended prey ($\lambda_2=\lambda_1$).

It is also instructive to plot $T_{\rm M}$ against the memory parameter
$\gamma$ (figure \ref{fig_transient-time}c).  The transient time $T_{\rm M}$
increases with forgetting rate $\gamma$ and diverges at a critical value,
denoted by $\gamma_{\rm min}$ in figure 3 of the main article.
The divergence indicates that asymptotic mutualism appears above
$\gamma=\gamma_{\rm min}$.

\subsection{Density-dependence}

We describe the effect of mimic density on the survival benefits.  We vary the
relative mimic density $\delta = n_2/n_1$, with fixed $n_1+n_2=1$.

With an infinite-memory predator ($\gamma=0$), the asymptotic attack
probabilities are simply the palatabilities $\lambda_1$ and $\lambda_2$ for
non-resembling prey ($r=0$), independent of the densities.  The $r=1$
asymptote   $\bar{\lambda} = (n_1 \lambda_1+n_2 \lambda_2) /
(n_1+n_2)$ depends on densities. 
The asymptotic mimic favorability 
\[
f_2 = \frac{\lambda_2}{\Bar{\lambda}} = \frac{\lambda_2 (1 + \delta)}{(\lambda_1 + \lambda_2
\delta)} \; 
\]
monotonously decreases with $\delta$ from $f_2(\delta \to
0)=\lambda_2/\lambda_1$ to $f_2(\delta \to \infty)=1$.  The density dependence
comes entirely from the denominator.

A finite predator memory ($\gamma\neq0$) makes the situation more complicated.
Now the $r=0$ asymptotes also depend on density, because the asymptotic attack
probabilities are determined by the competition between learning and
forgetting.  The learning pulls the attack probabilities $P_i$ towards the
palatibilities $\lambda_i$, while the forgetting tries to push them back to
the naive value $P_0$.
With increasing relative density $\delta$, the learning process becomes more
important compared to the forgetting process, which is density-independent, so
that the asymptotic $P_i$ get closer to the palatibilities $\lambda_i$.
For palatable mimics ($\lambda_2>0.5$), the $P_2^{(r=0)}(t\rightarrow\infty)$
monotonically increases with relative mimic density $\delta$.  The finite-$r$
asymptote, $P_2^{(r>0)}$, also grows with $\delta$, but at a different rate,
and as a result the ratio (favorability) can depend non-monotonically on
$\delta$, as illustrated in the main text.

\subsection{Further applications and extensions}

We outline below some further applications of our formulation.

\begin{figure}
\begin{center}
\centering
\includegraphics*[width=0.98\columnwidth]{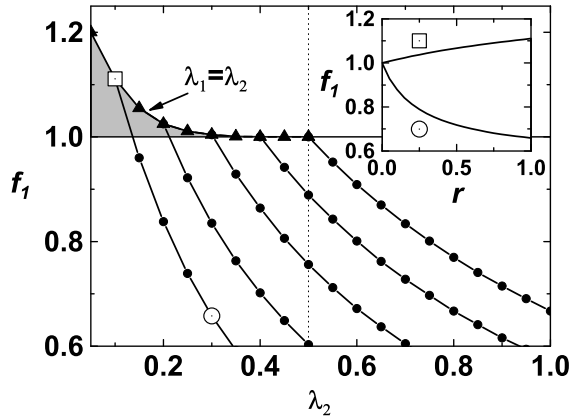}
\caption{
Search for mutualism with palatable mimic.
\\ Asymptotic model favorability $f_1^{(r=1)}$ as a function of mimic
palatability $\lambda_2$, obtained with cubic forgetting term $\gamma_{\rm c}
(P_0-P_{1,2})^3$, where $\gamma_{\rm c}=5 \times 10^{-4}$.
The sequence of $f_1$-vs-$\lambda_2$ curves (filled circles) each correspond
to a different value of $\lambda_1$, ranging from 0.1 for the leftmost curve
to 0.5 for the rightmost.
The tops of these curves (filled triangles) correspond to equally defended
prey, $\lambda_1=\lambda_2$.
Shaded area indicates the region of mutualistic mimicry ($f_1>1$) --- clearly
confined to $\lambda_2<0.5$, thus yielding a negative answer to the
``mutualism with palatable mimic'' question.
%
% Dotted vertical line shows the neutral palatability $\lambda_2=0.5$.
%
\\ Inset shows, for one mutualistic ($f_1>1$) and one parasitic ($f_1<1$)
case, that the dependence on resemblance coefficient $r$ is monotonic, so that
the same negative answer holds for arbitrary $r$.
}\label{fig_no-palatable-Muellerian}
\end{center}
\end{figure}

\subsubsection{Mutualistic mimicry with palatable mimic?}

The ease of calculations with the differential-equation setup encourages us to
ask more challenging questions, \emph{e.g.}, is there a situation in which a
\emph{palatable} mimic can be favorable for a defended model?  Within our
formulation, this means asking whether one can have benefit to the model
($f_1>1$) when $\lambda_1<P_0<\lambda_2$.

For both perfect ($r=1$) and imperfect ($0<r<1$) mimicry, our exact asymptotic
solutions [of equations (\textbf{3},\textbf{4}) in the main text] give a
negative answer for the above question: mutualistic mimicry requires
$\lambda_2<P_0$. 

To explore the question further, we have studied several modifications of our
basic equations, in particular
\\ 
\textbf{(A)} using a palatability-dependent learning coefficient,
$\alpha(\lambda) = \alpha_0(0.5+|\lambda-0.5|)$, as in Ref.~\cite{speed93}; 
\\
\textbf{(B)} replacing the linear forgetting term in equations
(\textbf{3},\textbf{4}) by various nonlinear
forms, such as $\gamma_{\rm c}(P_0-P)^3$ or $\tilde{\gamma}(P_0^2-P^2)$.
\\ None of these variants of predator behavior allowed for mutualistic mimicry
when $\lambda_1<P_0<\lambda_2$.
Although the negative statement is impossible to prove strictly, our results
indicate that mutualistic mimicry generically does not occur for palatable
mimic species, at least within a wide class of approximations formulated
within the framework represented in figure 1 in the main article.
In figure \ref{fig_no-palatable-Muellerian}, we demonstrate the absence of
mutualism for one of the variants we have treated.

\subsubsection{Resemblance and discrimination}

Our last example application is motivated by the fact that mimicry is rarely
perfect in nature.  The evolution of imperfect resemblance and the result of
predator discrimination errors are important active topics in evolutionary
theory \cite{MacDougallDawkins_AnimBehav98,
GavriletsHastings_coevolutionary_JTheorBiol1998,
Edmunds_imperfect-mimiry_Linnean00, Johnstone_imperfect-mimiry_Nature02,
Sherratt_imperfect-mimiry_BehavEcol2002}.  In our formalism, the ability of
the predator to discriminate between prey species (or the degree of
resemblance) is parametrized by a continuous variable $r$, providing the
opportunity to explore the whole resemblance spectrum.
For example, figures 2b, 2e, 3b, 3d plot attack probabilities as a function of
the resemblance spectrum.  

The evolution of resemblance can be thought of as modification of the $r$
variable.  We can interpret the $r$-dependence of the attack probabilities:
$dP/dr$ measures the driving force for evolutionary change.  We can treat
evolutionary dynamics by allowing $r$ to be a dynamical variable itself.  The
simplest equation would be of the type
\[
\frac{dr}{dt} ~=~ -\epsilon_1 \frac{\partial P_1}{\partial r} ~- \epsilon_2
\frac{\partial P_2}{\partial r}
\]
Our formalism is thus potentially well-suited for a simple mathematical
description of the dynamics of resemblance evolution.

\begin{acknowledgments}
We thank G.~Del~Magno, D.~ O~Maoileidigh, E.~M.~Nicola, and R.~Vilela for
commenting on the manuscript.
\end{acknowledgments}

\end{document}